\documentstyle[multicol,pre,aps,epsf]{revtex}
\begin{document}

\date{\today}
\title{Dynamical Scaling: the Two-Dimensional XY Model Following a Quench}
\author{F. Rojas$^{a,b*}$ and A. D. Rutenberg$^c$}
\address{$^a$Department of Theoretical Physics, University of Manchester, 
Manchester, M13 9PL, UK}
\address{$^b$\footnote{Current and permanent address.}
Centro de Ciencias de la Materia Condensada, UNAM,  
Apartado Postal 2681, Ensenada, B.C., Mexico 22800 }
\address{$^c$Centre for the Physics of Materials, Physics Department,
McGill University, Montr\'{e}al QC, Canada H3A 2T8}
\maketitle

\begin{abstract}
To sensitively test scaling in the 2D XY model 
quenched from high-temperatures into the ordered phase, we study
the difference between measured correlations and the (scaling) results of a
Gaussian-closure approximation. 
We also directly compare various length-scales. 
All of our results are consistent with dynamical scaling
and an asymptotic growth law $L \sim (t/\ln[t/t_0])^{1/2}$, though with a 
time-scale $t_0$ that depends on the length-scale in question. We then 
reconstruct correlations from the minimal-energy configuration consistent
with the vortex positions, and find them significantly different from
the ``natural'' correlations --- though both scale with $L$. 
This indicates that both topological
(vortex) and non-topological (``spin-wave'') contributions to correlations
are relevant arbitrarily late after the quench.  
We also present a consistent definition of dynamical
scaling applicable more generally, and emphasize how to generalize our
approach to other quenched systems where dynamical scaling is in question.
Our approach directly applies to planar liquid-crystal systems. 
\end{abstract}
\pacs{05.70.Ln,64.60.Cn,61.30.Jf}
%%%%%%%%%%%%%%%%%%%%%%%%%%%%%%%%%%%%%%%%%%%%%%%%%%%%%%%%%%%%%%%%%%%
\begin{multicols}{2}
\section{Introduction}

The study of non-equilibrium dynamics in systems with 
continuous symmetries has burgeoned \cite{Bray94}.  
Liquid-crystalline systems 
\cite{Shiwaku87,Nagaya92,Pargellis92,Pargellis94,Yurke92,Wong92,Wong93}, 
evolving after being quenched into 
an ordered phase, provide picturesque examples of topological defects and
their interactions.  Evolving systems of topological defects are also found
in applications from cosmology \cite{Turok97} to 
quantum Hall ferromagnets \cite{Rutenberg98,Rutenberg97}.

A relatively simple system with a continuous symmetry is the
two-dimensional XY ferromagnet with no disorder, 
which supports singular vortices that carry
topological charge and have logarithmic
interactions.  The equilibrium properties have spawned a rich and fertile 
literature punctuated by the work of 
Kosterlitz and Thouless \cite{Kosterlitz73}. 
More recently, the 
non-equilibrium behavior of the 2D XY model following a quench to below 
the Kosterlitz-Thouless critical temperature, $T_{KT}$, has been studied
theoretically 
\cite{Mondello90,Bray90,Blundell93,Bray93i,Yurke93,Blundell94,Lee95,Toyoki93}
and also experimentally \cite{Nagaya92,Pargellis94} with specially prepared 
liquid-crystal systems. Related 2D liquid-crystal systems have also
been studied theoretically \cite{Blundell92,Zapotocky95,Fukuda98} 
and experimentally \cite{Shiwaku87,Pargellis92,Wong93}. 

Following a quench at $t=0$ from a disordered phase into an ordered phase, 
a crucial issue is whether there is dynamical scaling 
\cite{Furukawa85} at late times $t$, where
\begin{equation}
\label{EQN:scaling}
	C(r,t) \equiv 
		\langle \vec{\phi}(x,t) \cdot \vec{\phi}(x+r,t) \rangle = f(r/L).
\end{equation}			
Here, $\vec{\phi}$ is the XY order parameter, $f(x)$ is a 
time-independent scaling-function for the two-point correlations, and
$L(t)$ is a growing length-scale that captures all of the correlation dynamics.
The explicit or implicit assumption of dynamical scaling underpins most
theoretical descriptions of phase-ordering structure
\cite{Bray94,Ohta82,Mazenko97,Bray94i}.  
Unfortunately, apart from a limited number
of solvable systems, there exist no theoretical approaches to {\em a priori} 
determine dynamical scaling.  Indeed, the presence or absence 
of dynamical-scaling remains an 
unresolved issue in the 2D XY model \cite{Blundell94,Zapotocky95}.  
This is surprising, since simple systems that break scaling are seen 
as exceptions \cite{multi}.
For example, the weak scaling violations in the conserved spherical 
model identified by Coniglio and Zannetti \cite{Coniglio89} are due to 
non-commuting spherical and asymptotic-time limits \cite{Bray92i} related to 
similar phenomena in equilibrium critical dynamics \cite{Ma74}.

Stronger scaling violations are found in
one- and two-dimensional systems with non-singular topological textures 
\cite{Rutenberg97,Rutenberg95i}.  These systems segregate into domains of 
similarly charged textures, similar to the morphologies 
seen in reaction-diffusion $A + B \rightarrow \emptyset$ systems 
\cite{Ovchinnikov78}.  The domain-size and the texture 
separation provide distinct growing length-scales.   Within this
context, the difficulty in resolving scaling in the 2D XY model can
be understood. Viewed as a plasma of overdamped 
charged vortices with logarithmic interactions \cite{mobility}, 
quenched from high-temperatures,
the 2D XY model sits exactly at the marginal dimension ($d=2$) 
below which segregated morphologies with strong scaling violations 
are expected, and above which a mixed morphology with only one length-scale, 
the particle separation, is seen \cite{Rutenberg98}.  Such {\em particle} 
systems are expected to 
scale, with no domain structure, at the marginal dimension
\cite{Rutenberg98}, however the asymptotic regime could be quite late.

With dissipative dynamics and the assumption of dynamical-scaling 
the predicted asymptotic 
growth-law of the characteristic length-scale is \cite{Bray94i}
\begin{equation}
\label{EQN:growth}
	L(t) \simeq  A (t/\ln[t/t_0])^{1/2},
\end{equation}
where $A$ and $t_0$ are the non-universal amplitude and time-scale,
respectively. 
This growth-law characterizes the correlations with a length 
$L_{1/2}(t)$, where $C(L_{1/2},t)=1/2$, as well as the vortex separation with
a length-scale $L_v(t)$, where the vortex density $\rho_{def} = 1/L_v^2$. 
These lengths will only differ by prefactors and by subdominant contributions
at late times. [Eqn.~\ref{EQN:growth} also describes the annihilation 
time of an isolated vortex-antivortex pair with 
an initial separation $L$ \cite{Yurke93}.]
The logarithmic factor is crucial, and stems from the logarithmic 
vortex mobility.  The same growth-law is
expected in liquid-crystal films with vortices \cite{Bray94i}.

The analytical evidence for scaling violations is mostly suggestive:
explicit violations in four-point correlations \cite{Bray93i} and multiple
energy-scales seen in energy-scaling calculations \cite{Bray94i}.  These 
would indicate multiple lengths which differ at most by logarithmic
factors, consistent with the marginal dimensionality within a reaction-diffusion
context \cite{Rutenberg98}.  Indeed, approximation schemes for correlation
functions in the 2D XY model typically find scaling but with no 
logarithmic factors (see, e.g., \cite{Mazenko97,Bray91}, see also
\cite{nolog}). Additionally, the 2D XY model 
quenched {\em between} two temperatures below $T_{KT}$, and 
coarse-grained to a fixed scale to eliminate bound vortex pairs, 
is solvable \cite{Rutenberg95} and 
dynamically scales without any logarithmic factor,  $L(t) \sim t^{1/2}$. 

Previous numerical evidence for scaling violations is stronger. 
Cell-dynamical simulations of XY models quenched to $T=0$ 
by Blundell and Bray
\cite{Blundell94} found that two-point correlations did not scale 
well with respect to the defect separation $L_v$, though they  scale 
with respect to the correlation length $L_{1/2}$ (see also
\cite{Blundell93,Toyoki93}). Mondello and Goldenfeld \cite{Mondello90} also
found indications of multiple length-scales. 
Simulations of nematic films by Zapotocky {\em et al.}
\cite{Zapotocky95} found a variety of effective growth exponents, though again
the correlation function appeared to scale (see also 
\cite{Blundell92,Fukuda98}). 
Other simulations on the 2D XY model at finite temperatures
have recovered the expected growth law \cite{Yurke93,Lee95}, 
and have found dynamical scaling \cite{Lee95}.  Simulations of quenches to
$T=0$ in hard-spin systems found dynamical
scaling of correlations even though the 
dynamics froze at late times \cite{Bray90}!

Experiments on liquid-crystal systems, following the pioneering
work by Shiwaku {\em et al.} \cite{Shiwaku87}, have recovered the 
$t^{1/2}$ growth of defect separation after a quench, though 
with insufficient resolution to determine logarithmic factors 
\cite{Nagaya92,Pargellis92,Pargellis94,Wong93} and with some difficulties in
achieving an unbiased (symmetric) quench \cite{Pargellis92,Pargellis94}. 
When measured, the structure \cite{Wong92,Wong93} and 
other two-point correlations are consistent with dynamical scaling
\cite{Nagaya92}.  

In this paper we want to clarify the existence or absence of 
dynamical scaling in the 2D XY model.  A successful strategy can then be 
applied more generally to systems that seem to violate scaling, in
particular to systems with more complicated collections of defects
\cite{multi}.

We first discuss the appropriate definition of dynamical scaling, within the
context of systems relaxing after a quench. We then derive approximate
forms for various correlation functions via Gaussian closure techniques, 
which impose scaling. While we do not expect them to exactly match the
measured correlations, they are used to normalize the measured values in
order to enhance our sensitivity to scaling or its absence.  In combination
with the growth-law, we have a ``null hypothesis'' which would be broken
by scaling violations. We
present our simulation data and find no evidence for scaling violations.
We then explicitly reconstruct a two-point gradient correlation function,
within the periodic system using only the vortex positions and charges, 
and find it significantly
different from the unreconstructed scaling form. However both correlations
scale with respect to the defect density. This indicates that both 
topological (vortex) and non-topological (``spin-wave'') contributions to the 
order parameter are asymptotically relevant, with characteristic lengths
that remain asymptotically proportional.

\section{Dynamical Scaling}
\label{SEC:scaling}

In phase-ordering, dynamical scaling colloquially means that there is a
single characteristic length scale growing in time. This leads to a
rough-and-ready symptom of dynamical scaling violations:
multiple length-scales with distinct growth-laws, see
for example \cite{Zapotocky95,Lee95i}.  While useful as a guide, this 
approach has limitations.  One must first identify each 
asymptotic growth law, i.e. the effective exponent
after it is constant in time and before finite-size effects of the sample
become important. Practically, at most one or two decades in time are available
in simulations if a $5\%$ exponent variation is tolerated, and 
often less than a decade in experiments. When the scaling prediction for the
growth-law is not {\em a priori} known, this approach on its own is dangerous. 
Indeed, sub-dominant corrections to the asymptotic growth law \cite{Bray98} 
can depend on the method used to extract the length-scale 
\cite{scalingcorrection}. Even the observation of two asymptotically distinct
length-scales does not demonstrate that they are dynamically
interconnected.  A silly example helps here: consider a sample made
from gluing together a conserved binary-alloy system (asymptotic
growth law $t^{1/3}$), 
and a non-conserved order-disorder alloy system (growth law $t^{1/2}$).
Clearly two-growth laws could be observed in the hybrid, 
but they should not imply scaling violations. 
[Such dynamically independent sub-systems would lead to correlation
functions that are sums of scaling functions.]
The situation is more complicated when both lengths
are observed within a homogeneous sample, such as the asymptotic 
behavior of monopoles and vortex lines in bulk nematics \cite{Yurke92}.
Non-trivial inter-relationships of observed lengths can generally
only be resolved with the help of simplified 
dynamical models, for example see \cite{Rutenberg95i,Siegert98}.  

A more precise definition of dynamical scaling is that two-point equal-time 
correlations have a time-independent scaling form,
see Eqn.~\ref{EQN:scaling}, which also implies scaling of the structure
factor
\begin{equation}
\label{EQN:structure}
		S(k,t) \equiv \langle \vec{\phi}(k,t) \cdot \vec{\phi}(-k,t) \rangle
				= L^d g(k L),
\end{equation}
where $g(x)$ is a time-independent scaling function.  This is
directly measured in scattering experiments, can be well
approximated analytically, and is easy to extract from simulations. For
systems with singular topological defects, such as domain walls, hedgehogs,
vortices, or vortex lines, a generalized Porods law \cite{Bray94} 
connects the density of defect core $\rho_{def}$ to the asymptotics
of the structure via 
\begin{equation}
\label{EQN:porod}
		S(k) \sim \rho_{def} k^{-(d+n)},     kL \gg 1, 
\end{equation}
where $n$ characterizes the defect type [for the 2D XY model, $n=d=2$]. 
This directly implies that the 
length derived from the defect density, $L_v$, is asymptotically
proportional to the
correlation length $L_{1/2}$ when the  correlations dynamically scale.

This definition is still incomplete, since systems can satisfy 
Eqn.~\ref{EQN:structure} yet have distinct lengths intimately connected 
by the dynamics --- e.g. in the 1D XY model \cite{Rutenberg95i}. 
Additionally, higher-point correlations can be constructed in the 2D XY model
which explicitly do {\em not} scale \cite{Bray93i,higherpoint}. Should these
be viewed as violations of dynamical scaling? 
Fortunately a self-contained definition of dynamical scaling 
exists, introduced by Bray and Rutenberg \cite{Bray94i}. 
In order to calculate the rate of free-energy dissipation in a
coarsening system, they additionally require 
the scaling of the time-derivative correlation function
\begin{eqnarray}
\label{EQN:Ti}
T(r,t) & \equiv & \left\langle \partial_t \vec{\phi}(x,t)
		\cdot \partial_t \vec{\phi}(x+r,t)\right\rangle 
			=  (\dot{L}/L)^2 F( r/L ) 
\label{EQN:Tii}
\end{eqnarray} 
where $F$ is a new time-independent scaling function and $\dot{L} \equiv dL/dt$.
Note that power-law growth, with or without additional logarithmic factors, 
implies that the prefactor $(\dot{L}/L)^2 \sim 1/t^2$.
If dynamic scaling holds both for $T(r,t)$, as just defined, 
and for $C(r,t)$, then the growth 
exponent can be determined through a self-consistent energy-scaling approach
\cite{Bray94i,generalscaling}.  This restricted definition of dynamical 
scaling, of both $C(r,t)$ {\em and} $T(r,t)$, picks up the scaling
violations of the 1D XY model \cite{Rutenberg95i}, and clearly separates the 
role of two-point from higher-point correlations \cite{higherpoint}.   We use
this restricted definition here, and recommend
it in the study of systems where dynamical scaling is questioned but 
Eqn.~\ref{EQN:scaling} seems to be satisfied. 

\section{Dynamics}

We study purely dissipative quenches of 2D XY models from well above 
to below the Kosterlitz-Thouless transition temperature $T_{KT}$. 
Because of the line of critical points in the 2D XY model \cite{Kosterlitz73}
the correlations in quenches to $0< T < T_{KT}$ have a modified
scaling form \cite{Lee95,Rutenberg95}. Essentially, 
critical equilibrium correlations have no characteristic length-scale and
so the standard coarse-graining \cite{Bray94} 
to make temperature irrelevant to large-scale correlations is impossible. 
However, there is no indication that temperature changes dynamical scaling, 
or its absence, in the 2D XY model. Accordingly, 
in this paper, we only investigate quenches
to $T=0$. The non-conserved coarse-grained dynamics \cite{conserved} are 
\begin{eqnarray}
\label{EQN:modeldyn}
	F[\vec{\phi}] &=& \int d^2 x \left[ (\nabla \vec{\phi})^2
	+V_0 (\vec{\phi}\,^2-1)^2 \right], 
	\nonumber \\
	\partial_t \vec{\phi} &=& - \Gamma \delta F /\delta \vec{\phi},
	\nonumber \\
	\langle \vec{\phi}({\bf x},0) \cdot \vec{\phi}({\bf x'},0) \rangle 
		&= & \Delta \delta({\bf x}-{\bf x'}),
\end{eqnarray}
where $\Gamma$ is a kinetic coefficient that sets the time-scale, 
$V_0$ is the potential strength that sets the `hardness' of the vector
spins, and $\Delta$ characterizes the initial disordered state. 
The orientation of the two-component order-parameter $\vec{\phi}({\bf x})$
defines an angle $\theta({\bf x}) \in [0,2\pi]$, which is identical to the XY
phase.  The numerical implementation of the dynamics 
is discussed below in Sec.~\ref{SEC:methods}.

\begin{figure}
\epsfxsize=3.5truein % size of blank space generated
\epsfysize=3.1truein % size of blank space generated 
\vbox{
\vskip 0.15truein %offset down from top of region
\epsfbox[95 180 550 650]{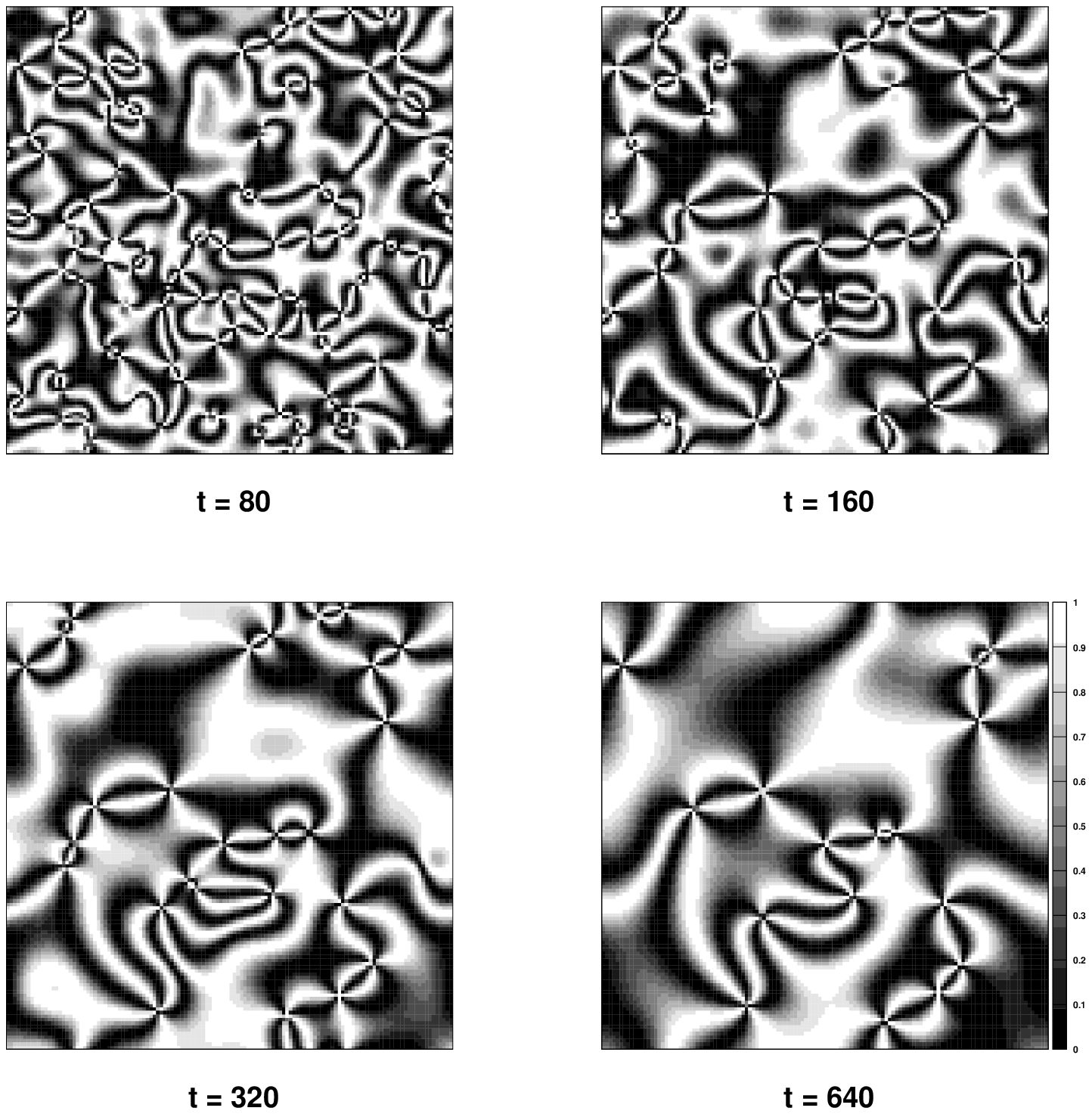}}
\caption{Schlieren patterns in various times after a quench
of the 2D XY model in a size $128 \times 128$ system.  
The intensity is $\sin^2(2\theta)$, where
$\theta$ is the local XY phase. Each vortex emanates $8$ brushes, 
alternating white and black.}
\label{FIG:schlieren}
\end{figure}

In overview of the evolution: 
we start with a random high-temperature configuration and quench to 
$T=0$.  The order parameter locally equilibrates, but 
competition between degenerate ground-states leads to topologically
stable vortices, with integer charges. The annihilation of oppositely
charged vortices drives the subsequent dynamics, and characterizes one 
possible growing length scale --- the vortex separation $L_v$.  
Of course, the order-parameter field around a moving vortex is not rigidly
comoving \cite{Bray94i}, and so 
non-singular ``spin-wave'' distortions are generated by the dynamics even
at $T=0$.  The dynamics, emphasizing the vortices, can 
be visualized with a Schlieren pattern, see Fig.~\ref{FIG:schlieren},
analogous to those used in the study of liquid-crystal films
\cite{liquidcrystal}.

\subsection{Scaling Correlations from Gaussian Closure}
\label{sec:xygaf}

Several approximation schemes eliminate high-order correlations in the
evolution equation for two-point correlations 
\cite{Bray94,Ohta82,Mazenko97,Bray91,Bray92}. We use a Gaussian-closure 
approximation, which gives quite good two-point correlations.  
We will use the results to
normalize our correlations. This allows for a more sensitive
test of scaling properties than has been possible before, and also
highlights weaknesses of this approach (see also \cite{Blundell93,Yeung94}).

For general $O(n)$ fields, we start with the
Bray-Humayun-Toyoki (BPT) approach \cite{Bray91}.
We introduce an auxiliary field $\vec{m} $ parallel to the order parameter, 
$\widehat{m} = \widehat{\phi}$.  The zeros of 
$\vec{m}$ match the positions of the topological defect cores, while 
$|\vec{m}|$ is roughly the distance to the closest defect core. 
Assuming a Gaussian probability distribution for $\vec{m}$
results in two-point correlations between $(r_1,t_1)$ and $(r_2,t_2)$:
\begin{equation}
\label{EQN:bpt}
	C_g(r,t_1,t_2) = \frac{n \gamma}{2 \pi}\left[
	B\left(\frac{1}{2},\frac{n+1}{2}\right)
	\right]^2F\left(\frac{1}{2},\frac{1}{2};\frac{n+2}{2};\gamma^2\right),
\end{equation}
where $r= |{\bf r}_2 - {\bf r}_1 |$, $B(x,y)$ is the beta function, and 
$F(a,b;c;z)$ is the hyper-geometric function. 
The result is expressed in terms of the the normalized two-point,
two-time correlation function of $\vec{m}$:
$\gamma  = \langle m(1)m(2) \rangle/ [\langle
	m^2(1)\rangle\langle m^2(2)\rangle ]^{1/2}$.

The various approximation schemes
schemes differ on the manner of determining $ \gamma $.
We use the the systematic approach introduced by Bray and Humayun
\cite{Bray92} which produces
\begin{equation}
\label{EQN:gamma}
	\gamma(r,t_1,t_2) = \left( \frac{4t_1t_2}{(t_1+t_2)^2} \right)^{d/4} 
	\exp( -r^2/[4(t_1+t_2)]), 
\end{equation}
where $d$ is the spatial dimension.
For equal-time correlations, we obtain the scaling form
$C_g(r,t) = f_{BPT}(x)$, where $x = r/L$ and $L(t) = (4t)^{1/2}$.
This highlights a problem with all existing correlation-closure approaches 
as applied to 2D XY models, since while they recover a scaling form 
they miss the logarithmic factor in the growth-law \cite{nolog}.
The same scaling variable is used in the time-derivative correlation function 
\begin{equation}
\label{EQN:tdc}
		T_g(r,t) = \frac{1}{16t^2} \left [ \gamma^2 x^4 C_{\gamma \gamma}(x) +
			\gamma ( x^4-4x^2 + 2d) C_{\gamma}(x) \right],
\end{equation}
where $C_\gamma \equiv \partial C_g / \partial \gamma$ and 
$C_{\gamma \gamma} \equiv \partial^2 C_g / \partial \gamma^2$. 

\section{Simulation}
\subsection{Simulation Methods} 
\label{SEC:methods}

We use a standard CDS update \cite{Oono87} for soft spins,
$\vec{\phi}({\bf i},t)$,  on a periodic lattice, 
where $t$ is now a discrete integer time and $\bf i$ is the position:
\begin{equation}
\label{EQN:cdsdynamics}
	\vec{\phi}({\bf i},t+1) = \frac{D}{4}
	\sum_{\bf j} \left[ \vec{\phi}({\bf j},t) - \vec{\phi}({\bf i},t) 
			\right] 
			+ E \, \, \widehat{ \! \!\phi}({\bf i},t) \tanh \left[ \left|
			\vec{\phi}({\bf i},t)\right| \right],
\end{equation}
where $\widehat{ \! \phi} =  \vec{\phi}/\left| \vec{\phi} \right|$ 
is the unit vector. We use the standard values $D=0.5$ and $E=1.3$.  
The dynamics are stable and have 
the same attractors as Eqn.~\ref{EQN:modeldyn}.
We do not observe pinning effects in quenches to $T=0$ (see also
\cite{Mondello90,Blundell93,Blundell94,Blundell92,Zapotocky95}).
The random initial conditions are chosen uniformly for each component
from $\left[ -0.1,0.1 \right ]$.

We identify vortices with three methods that prove equally
effective: by looking for
the zeros in the vector field, by looking for plaquettes around which the 
phase rotates through $\pm 2\pi$,
and by finding the peaks on the local energy density 
$ E_{\bf i} = -\sum_{\bf j} \vec{\phi}({\bf i}) \cdot \vec{\phi}({\bf j})$,
where the sum is over nearest neighbors of site ${\bf i}$. Due
to the periodic boundary conditions, the system has no net vorticity.

In addition to tracking the number of vortices, we 
measure several correlations of the ``hardened'' order parameter,
$\widehat {\! \phi}({\bf j},t)$:
\begin{equation}
\label{EQN:simcor1}
	C(r,t) = \left\langle \widehat {\! \phi}({\bf j},t) 
	\cdot \widehat{ \! \phi}({\bf j}+ {\bf r},t) \right\rangle.
\end{equation} 
The average $\langle \dots \rangle$ is over 
the independent sets of initial conditions, and includes a spherical
average and an average over lattice sites ${\bf j}$. 
The structure factor is also calculated:
\begin{equation}
	S(k,t) = \left\langle \vec{\phi}({\bf -k},t) \cdot
	\vec{\phi}({\bf k},t)\right\rangle.
\end{equation}
We also measure the time derivative correlation function, 
\begin{equation}
\label{EQN:tdcor}
	T(r,t) = \left\langle \delta_t \vec{\phi}({\bf j}) 
			\cdot \delta_t \vec{\phi}({\bf j}+ {\bf r}) \right\rangle ,
\end{equation}
where $\delta_t \vec{\phi} = \vec{\phi}(t+1) - \vec{\phi}(t)$
is a finite difference approximation for the time derivative.

To probe the distinction between vortex and non-vortex contributions to 
correlations, we measure a phase-gradient correlation function:
\begin{eqnarray}
\label{EQN:gradtheta}
	D(r,t) &\equiv& \langle {\bf \nabla} \theta({\bf j} + {\bf r},t) 
								 {\bf \nabla} \theta({\bf j},t) \rangle, \\
\label{EQN:gradthetascale}
		&=& h(r/L)/L^2,
\end{eqnarray}
where the second line is the natural scaling ansatz for the correlations.
Note that $\langle {\bf \nabla} \theta \rangle =0$.  
We then reconstruct the vortex contribution $D_r(r,t)$ directly
from the charges and locations of the vortices at a given time. 
From the vortex positions we build up the 
phase field ${\bf \nabla} \tilde{\theta}({\bf j})$ 
using the periodic image of the 
minimal energy solution for each single vortex, $ \nabla^2\tilde{\theta} = 0$,
due to Gr{\o}nbech-Jensen \cite{Gronbech96}:
\begin{eqnarray}
\label{EQN:gronbech}
	d\theta/dx &=& - \pi \sum_{n=-\infty}^\infty \sin(2\pi y)/
								[ \cosh(2 \pi(x+n))-\cos(2 \pi y) ], 
								\nonumber \\
	d\theta/dy &=&  \pi \sum_{n=-\infty}^\infty \sin(2\pi x)/
								[ \cosh(2 \pi(y+n))-\cos(2 \pi x) ],
								\nonumber
\end{eqnarray}
where $x$ and $y$ are the relative coordinates of the vortex in a system of size
unity.  The solutions for every vortex (with $\pm 1$ factors for vortices
and anti-vortices, respectively) were added together for every point in the
system to obtain the fully-periodic {\em minimal-energy}
phase-field consistent with the vortex configuration.
[Direct reconstruction of the order-parameter field  $\vec{\phi}$
proved intractable due various counter-charge effects 
imposed by the periodic boundary conditions. In principle we could use
our $\nabla \theta$ reconstruction to recover the order-parameter field
with additional line-integrations.]
To obtain more accurate vortex positions, we first identify the
lattice plaquette by windings or energy peaks, then we use bilinear 
interpolation \cite{Press94} to more accurately 
locate the zero of the order parameter within 
the plaquette. The sign of the vortex is determined by the 
winding of the phase field around the plaquette.

\subsection{Simulation Results} 
\label{SEC:results}

We simulate a size $512\times 512$ system, averaging over 
$40$ independent samples. We check
that there are no significant finite size effects in comparison to a
$256\times 256$ size system, with $20$ samples. The data for reconstructed
correlations is currently restricted to the $256 \times 256$ size system.

\begin{figure}
\epsfxsize=3.5truein
\epsfysize=5.0truein
\vbox{\vskip 0.15truein
\epsfbox[10 0 460 750]{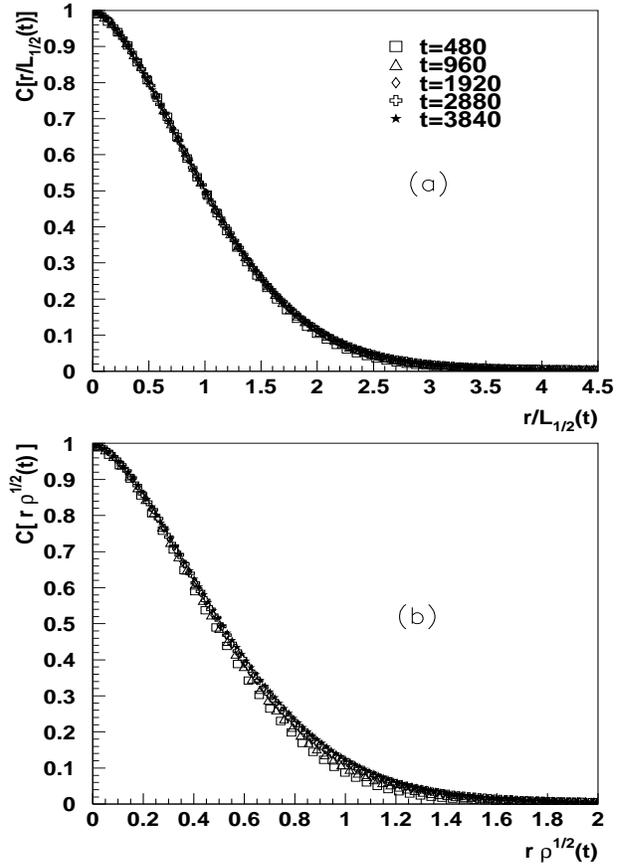}}
\caption{a) Real-space correlations vs $x=r/L_{1/2}$, where
$C(L_{1/2}) = 1/2$. The continuous
curve represents the theoretical prediction, $f_{BPT}(x)$. 
b) Attempted scaling with respect to the vortex density.}
\label{FIG:corrxy}
\end{figure}

In Fig.~\ref{FIG:corrxy} a), we plot $C(r,t)$ with respect to distance scaled
by $L_{1/2}$ [$C(L_{1/2},t)=1/2$]. The scaling is excellent, and the
Gaussian-closure result ($f_{BPT}$, solid line) is indistinguishable from
the data.  In Fig.~\ref{FIG:corrxy} b), however, the scaling collapse is not
good with respect to the vortex separation [$L_v$, where $\rho = 1/L_v^2$]
\cite{Blundell94}. It is difficult to tell from this second 
plot alone whether scaling simply has a later onset time, or if scaling 
violations are indicated.  This must be determined by a direct comparison of
the length-scales $L_v$ and $L_{1/2}$, as well as by a study of the
time-derivative correlations $T(r,t)$, as discussed in Sec.~\ref{SEC:scaling}.

\begin{figure}
\epsfxsize=3.5truein
\epsfysize=2.7truein
\vbox{\vskip 0.15truein
\epsfbox[40 0 450 328]{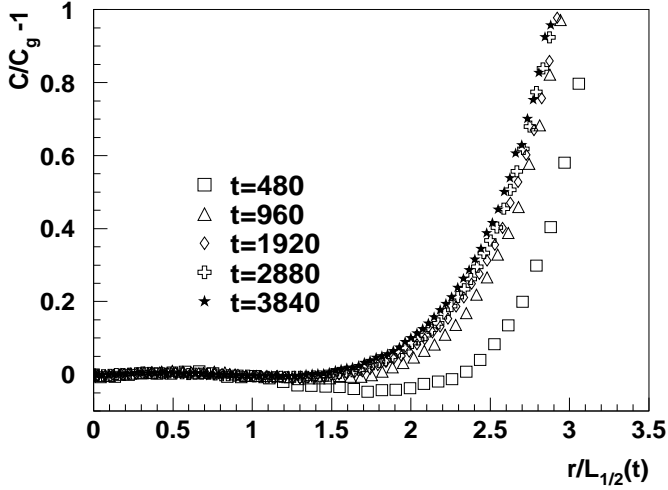}}
\caption{The difference between measured correlations, $C(r,t)$, and
the Gaussian-closure prediction, $C_g$, normalized by $C_g$ and plotted
against scaled distance.} 
\label{FIG:cdivgauss}
\end{figure}

By normalizing the correlations with the Gaussian-closure result, $C_g$, we
can sensitively probe scaling with the real-space correlations, see
Fig.~\ref{FIG:cdivgauss}. While $C_g$ is clearly too small at large 
scaled distance, correlations scale relatively well for $t \gtrsim 1000$. 

\begin{figure}
\epsfxsize=3.5truein
\epsfysize=2.7truein
\vbox{ \vskip 0.15truein
\epsfbox[20 0 450 330]{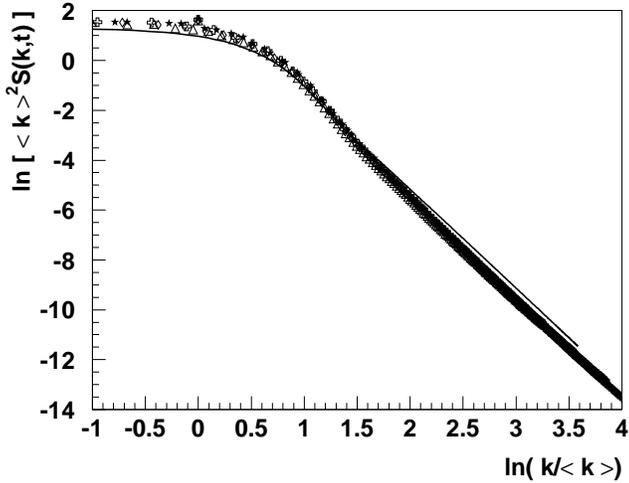}}
\caption{The structure factor in a log-log Porod plot. The
first moment, $\langle k \rangle$, is used to rescale momenta. 
The continuous line is the Gaussian-closure prediction. Symbols are the same
as the previous figure.}
\label{FIG:strucxy}
\end{figure}

The structure factor scales with respect to $L_k \equiv 1/\langle k
\rangle$, its inverse first moment, see Fig.~\ref{FIG:strucxy}. Also shown 
(solid line) is the Fourier transform of the Gaussian-closure prediction, which
slightly but systematically under and over-estimates the structure. By using
a  log-log plot we emphasize the $S(k) \sim \rho_{def} k^{-4}$ generalized
Porod tail for $k/ \langle k \rangle \gtrsim 2$, as per Eqn.~\ref{EQN:porod}.
The good scaling of the Porod tail, which is determined by the vortex 
density, indicates that $L_k \sim L_v$ asymptotically. 

\begin{figure}
\epsfxsize=3.5truein
\epsfysize=2.7truein
\vbox{\vskip 0.15truein
\epsfbox[0 0 470 550]{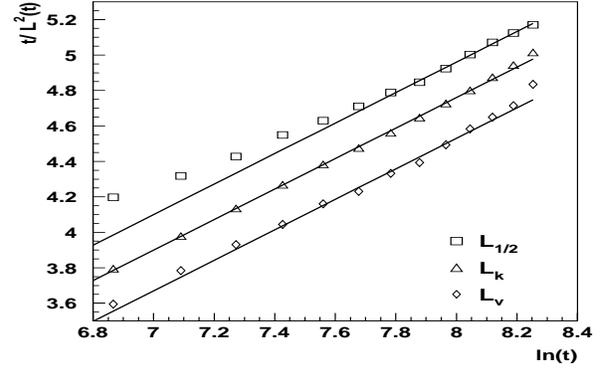}}
\caption{We plot $t/L^2$ vs $\ln{t}$ for three lengths: $L_{1/2}$, $L_k$,
and $L_v$. The observed linear dependence at late times indicates that the 
dynamical scaling growth law, Eqn.~\protect\ref{EQN:growth}, holds. As shown
by the parallel straight-lines, the offset (given by $t_0$) is non-universal.}
\label{FIG:defxy}
\end{figure}

We now directly test the assumption that all lengths asymptotically have the
scaling growth-law of Eqn.~\ref{EQN:growth} by plotting $t/L^2$ vs $\ln{t}$
for $L_{1/2}$, $L_k$, and $L_v$, in Fig.~\ref{FIG:defxy}. The scaling
prediction is a linear plot, with non-universal slope and intercept given by
the amplitude $A$ and time-scale $t_0$. [Both of these can vary from one
length-scale to another.] Linearity is observed for $\ln{t} \gtrsim 7.7$
($t \gtrsim 2200$), in agreement with Fig.~\ref{FIG:cdivgauss}.  We have 
fit them with straight lines with the same amplitude $A$ but different
$t_0$.  The correlation length $L_{1/2}$ has the strongest corrections to
scaling, which is one cause of the bad scaling of $C(r,t)$ when plotted vs.
$L_v$ in Fig.~\ref{FIG:corrxy} b).
It is worth noting that the growing length scales can also be well 
fit using 
effective exponents of $0.42$, $0.40$, and $0.40$ ($\pm 0.01$),
respectively, without logarithmic factors --- see also 
\cite{Blundell94,Zapotocky95}. However, if 
these effective exponents were asymptotically valid,
and hence disagreed with the scaling prediction of Eqn.~\ref{EQN:growth},
we would not see scaling in the correlations \cite{Bray94i}. 

\begin{figure}
\epsfxsize=3.5truein
\epsfysize=3.25truein
\vbox{ \vskip 0.15truein
\epsfbox[0 0 470 440]{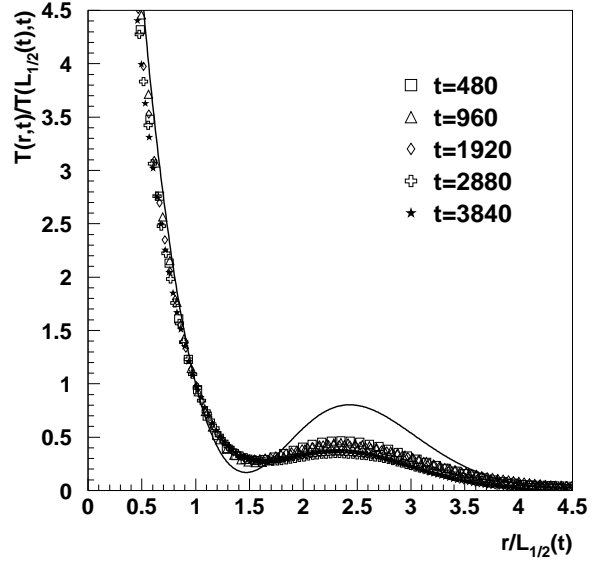}}
\caption{Scaling plot of $T(r,t)/T(L_{1/2},t)$ vs $r/L_{1/2}$. The continuous
curve is the theoretical prediction of the Gaussian
closure scheme --- significant deviations are apparent.}
\label{FIG:tdc}
\end{figure}

While the two-point correlations $C(r,t)$ and $S(k,t)$ support dynamical
scaling, we must also investigate the time-derivative correlation function,
$T(r,t)$, as discussed in Sec.~\ref{SEC:scaling}. In Fig.~\ref{FIG:tdc}, we
scale lengths with respect to $L_{1/2}$, and remove the prefactor in 
Eqn.~\ref{EQN:Tii} by plotting $T(r,t)/T(L_{1/2},t)$. While scaling only
sets in for $t \gtrsim 2000$, it is supported by the data. 
This correlation function has much more structure than the equal 
time correlations, such as a local maximum at $x \approx 2.3$ and 
a logarithmic divergence at small-$x$ due to fast vortex annihilations.
As a result, it provides a more stringent test of the 
Gaussian-closure approximation. We find significant discrepancies, the first 
to be found in two-point correlation functions.  

\begin{figure}
\epsfxsize=3.5truein
\epsfysize=2.0truein
\vbox{\vskip 0.15truein
\epsfbox[0 0 440 240]{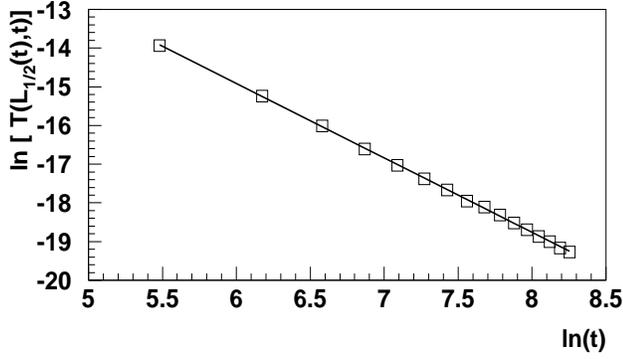}}
\caption{Log-log plot of the time dependent prefactor of 
the time-derivative correlation function. The
best fit over the range shown 
yields a decay $1/t^\mu$ with an exponent $\mu= 1.96$. Varying
the fit range yields $\mu= 2.0 \pm 0.1$.}
\label{FIG:ptdc}
\end{figure}

Further confirmation of scaling in $T(r,t)$ is found by exploring the 
time dependence of the amplitude $ T(L_{1/2},t) \sim
t^{-\mu}$, see Fig.~\ref{FIG:ptdc}. The
scaling form in Eqn.\ref{EQN:Tii} gives $\mu=2$ (independent of 
logarithms), and we find $\mu= 2.0 \pm 0.1$. This is consistent with scaling.
In combination with the scaling of $C(r,t)$ [and $S(k,t)$], and the
consistency of the growth laws of {\em all} measured length scales with the
scaling result, we conclude that the quenched 2D XY model dynamically scales.

\begin{figure}
\epsfxsize=3.5truein
\epsfysize=3.5truein
\vbox{\vskip 0.15truein
\epsfbox[34 0 440 440] {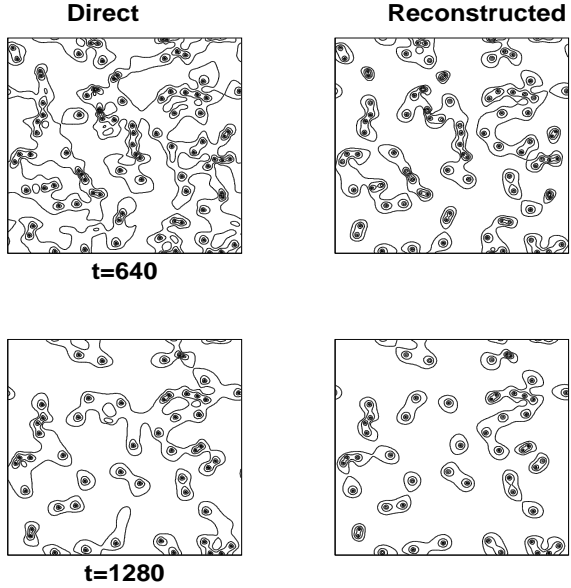}}
\caption{ Snapshots of $|\nabla \theta|$ for a $256 \times 256$ size system
at two times after the quench. The left column shows the direct
$|\nabla \theta|$ , with contour levels at $0.1$, $0.2$, $0.4$, $0.8$, and
$1.6$. [Note that the lattice spacing defines a unit of length, so the
largest gradient magnitude is $\pi$.] The right column shows 
$|\nabla \theta|$ periodically reconstructed using only the vortex
positions, with the same contour levels.
Significant differences between the direct and reconstructed $\nabla \theta$
field can be seen away from the immediate vicinity of the vortices.}
\label{FIG:gradtheta}
\end{figure}

\begin{figure}
\epsfxsize=3.5truein
\epsfysize=4.0truein
\vbox{ \vskip 0.15truein
\epsfbox[0 0 440 525]{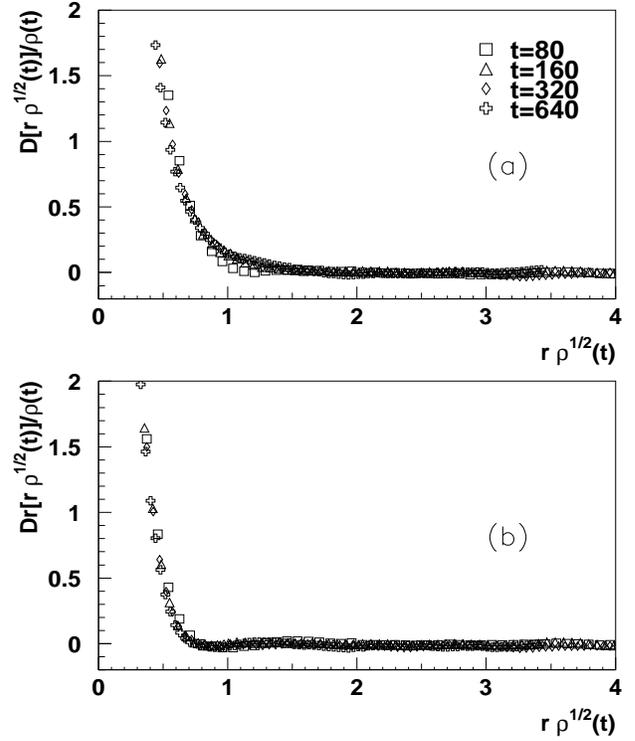}}
\caption{Direct and reconstructed $\nabla \theta$ correlations, $D(r,t)$, and
$D_r(r,t)$, respectively, normalized by vortex density for a scaling plot
vs scaled distance. Scaling is observed after the earliest time.} 
\label{FIG:gradre}
\end{figure}

In the equilibrium 2D XY model, the singular (vortex) and 
non-singular (spin-wave) degrees of freedom have independent contributions
to the free energy \cite{Chaikin95}.  
Could it be possible for such distinct
`singular' and `non-singular' length-scales to exist in phase-ordering 
systems (see, e.g., \cite{Kramer94})? 
If separation of vortices and spin-waves occurs, we expect
spin-wave contributions to have a characteristic scale $L \sim t^{1/2}$,
i.e. to have a faster decay with no logarithmic factor \cite{Bray94i}. In 
which case, the direct correlations should either have scaling violations due 
to the different length scale or the spin-waves should be asymptotically 
irrelevant --- leaving the direct and reconstructed correlations 
asymptotically equal at late times. As can be seen from the snapshots of
$|\nabla \theta|$ in Fig.~\ref{FIG:gradtheta}, 
the reconstruction maintains the vortex locations and is periodic. 
Indeed, the reconstruction provides the 
{\em minimal energy} configuration consistent with vortex positions  --- in
other words any `non-singular' contribution is absent.
In Fig.~\ref{FIG:gradre}, the correlation function for the direct
and reconstructed fields are shown as a function of the scaled distance.
We first notice that both correlations scale with respect to $L_v =
\rho^{-1/2}$ but with different functional forms. $D_r$ has a sharper knee
at $r \rho^{1/2} \simeq 0.7$, for example. This knee reflects the faster
decay of $\nabla \theta$ from the vortex core in the reconstructed 
configurations, as is apparent in Fig.~\ref{FIG:gradtheta}.
The significant 
differences between the bare and reconstructed correlations in the scaling
limit indicate that both vortex
and ``spin-wave'' contributions are relevant to the direct correlations, and
that the separation seen in static properties does not hold in the dynamics.

\begin{figure}
\epsfxsize=3.5truein
\epsfysize=4.0truein
%\vbox{\hskip 0.25truein \vskip 0.15truein
\vbox{ \vskip 0.15truein
\epsfbox[0 0 440 525]{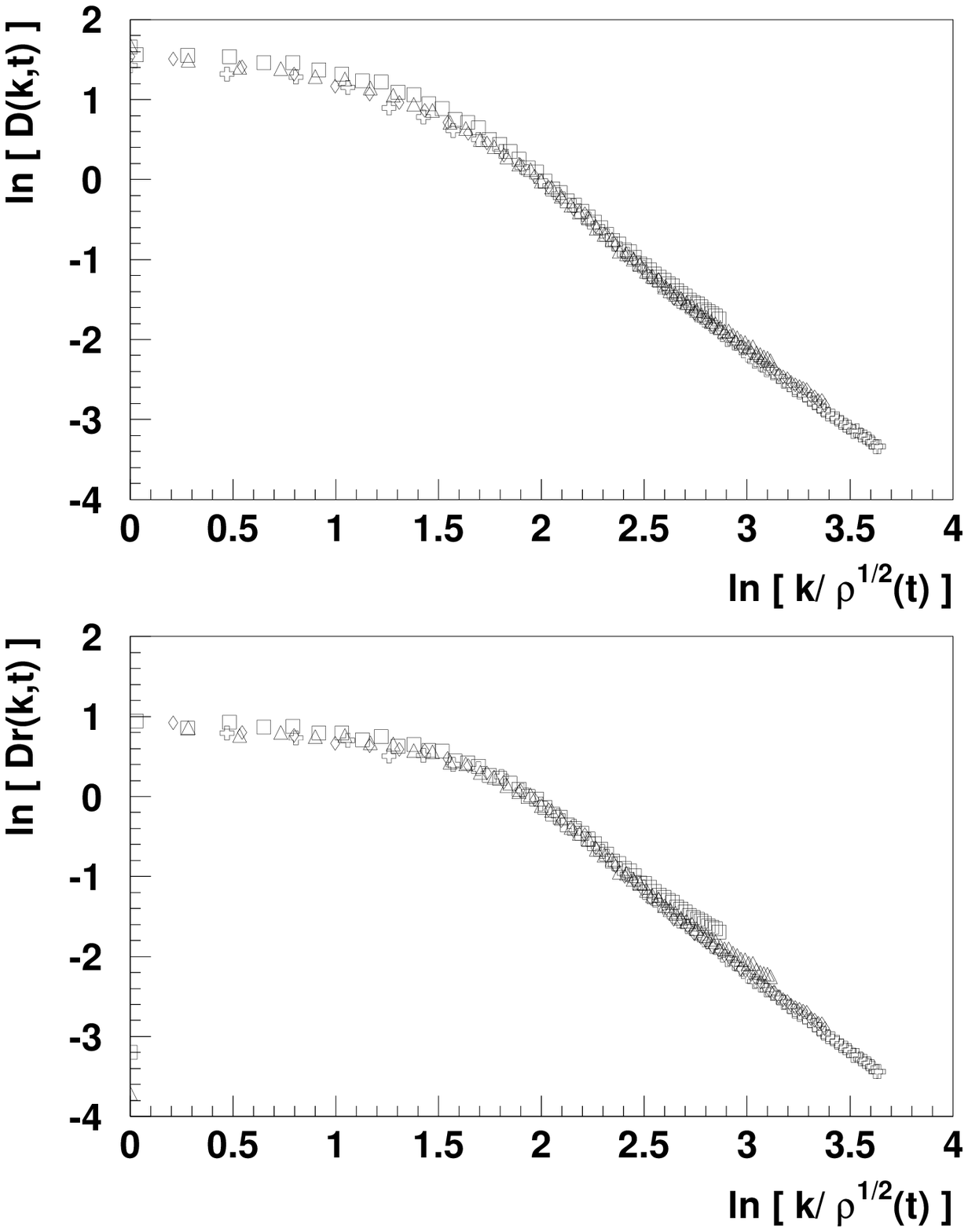}}
\caption{Porod plots of Fourier-transformed 
direct and reconstructed $\nabla \theta$ correlations, $D(k,t)$ and
$D_r(k,t)$, respectively. Symbols are the same as the previous figure.} 
\label{FIG:porodgradre}
\end{figure}

The Porod plot of the Fourier-transformed correlations, see
Fig.~\ref{FIG:porodgradre}, further highlights the differences (note the
$k \rightarrow 0^+$ intercepts). It is interesting that while 
$\langle \nabla \theta \rangle =0$, the scaling curve has a non-conserved
character. This is similar to correlations in globally-conserved systems.
We also observe a $k^{-2}$ Porod tail for $k/\rho^{1/2} \gtrsim 2$, which 
is expected from Fourier transforming the real-space scaling ansatz,
Eqn.~\ref{EQN:gradthetascale},
and setting the amplitude of the Porod tail proportional to
the vortex density $1/L^2$.  The Porod tail has the same amplitude between
the direct and reconstructed correlations, reflecting the singular structure
of the vortex cores. 

\section{Summary and Discussion}
We find {\em no evidence for scaling violations} in the 2D XY model. 
All lengths, $L_{1/2}$, $L_v$, and $L_k$, have the same asymptotic form given 
by Eqn.~\ref{EQN:growth}, albeit with different non-universal coefficients. 
Real-space correlations, structure, and time-derivative correlations all
scale as expected.  Phase-gradient correlations, reconstructed from the 
vortex locations to have minimal energy and hence no spin-wave
contributions, differed significantly from the direct correlations,
indicating that both vortex and non-singular ``spin-wave'' contributions
are asymptotically relevant.  We expect similar results to hold in
closely related planar liquid-crystal systems.

We have also shown how Gaussian-closure approximations can
be useful to sensitively explore scaling. This has the added benefit
of testing the approximation schemes. In particular we find significant
discrepancies with respect to 
the measured time-derivative correlations, $T(r,t)$.
More generally, we emphasize the role
of sensitive null-like tests in checking apparent scaling violations. For 
example we plot the length-scales vs the expected growth law so that linear 
behavior is expected if scaling is obeyed.  When scaling predictions are
available, and in the face of transient corrections to scaling, this is 
preferable to the measurement of effective exponents.

One can never absolutely rule out scaling violations, if only because
simulations and experiments can never reach $t \rightarrow \infty$. Each
length in a system that dynamically scales will generically have different 
corrections to scaling. In quenches of the 2D XY model
the leading correction is described well by $t_0$, the time-scale of the 
logarithmic factor.  Since scaling violations seem to be rare in quenched
systems, the assumption should be that systems dynamically scale
without strong evidence to the contrary --- including the inability to 
perform a scaling collapse with {\em any} length-scale for either $C(r,t)$
or $T(r,t)$.  This provides a self-consistent confirmation of dynamical
scaling, provided the lengths used for the collapse are consistent with 
the same asymptotic growth-law.  For some systems, including this one, 
the scaling growth-law can be independently determined.  This is invaluable
when long-lived (logarithmically decaying) corrections to scaling are expected. 
The scaling of some other lengths in the 
problem can sometimes also be required for consistency. In this case
the defect-separation scale $L_v$ is needed to set the Porod amplitude, 
and hence must be consistent with the lengths $L_k$ and $L_{1/2}$ 
extracted from $S(k,t)$ and $C(r,t)$, respectively.  

\section{ Acknowledgments} 
F. Rojas thanks CONACYT (Mexico) and EPSRC (UK) grant GR/J24782, while
A. D. Rutenberg thanks the NSERC, and {\it le Fonds pour la Formation
de Chercheurs et l'Aide \`a la Recherche du Qu\'ebec}.  We would like to thank 
Alan Bray, Rob Wickham, and Martin Zapotocky for useful discussions.

\pagebreak 

\pagebreak 

\pagebreak

\end{multicols}

\begin{thebibliography}{99}
\bibitem{multi} For multiple defect species, for instance with a $O(2) \times
Z_2$ ground-state degeneracy \cite{Lee95i} or for $d=3$ nematic quenches 
\cite{Yurke92}, multiple length-scales may be common. It remains to be seen
if these length-scales are coupled through the dynamics, i.e. if scaling
violations occur.
\bibitem{mobility} The analogy with an interacting particle system
is not exact since vortices in the 2D XY model have effective mobilities 
that depend on the local structure (see discussion in \cite{Rutenberg98}).
\bibitem{scalingcorrection} While the exponent of the leading correction
to a growth law is probably universal (between, for example, $L_{1/2}$ and 
$L_v$), a non-universal shift in the amplitude ratio between the growth 
law and the leading correction is sufficient to shift the effective exponent
at finite times.
\bibitem{higherpoint} Higher-point correlations that cannot be
expressed in a simple scaling form \cite{Bray93i}
are subdominant, in that they only probe the vicinity of defects and 
thus have negligible weight compared to two-point correlations.  It may
be that multiple length-scales can be found in any system with singular
defects if sufficiently high-order correlations, restricted to the vicinity
of the defects, are considered.
\bibitem{generalscaling} One could argue 
that a scaling form for full two-time correlations,
$C(r,t,t') \equiv \langle \vec{\phi}(x+r,t) \cdot \vec{\phi}(x,t') \rangle$,
is the most natural since it would include autocorrelations. However this
two-time correlation is
impractical to measure and is 
not necessary to determine the scaling growth-law. 
We use a less restrictive definition
of dynamical scaling, using only $C(r,t)$ and $T(r,t)$.
\end{thebibliography}
\end{document}